\documentclass[12pt]{article}
\usepackage{amssymb,graphicx}
\newcommand{\be}{\begin{equation}}
\newcommand{\ee}{\end{equation}}
\newcommand{\bea}{\begin{eqnarray}}
\newcommand{\eea}{\end{eqnarray}}
\newcommand{\beas}{\begin{eqnarray*}}
\newcommand{\eeas}{\end{eqnarray*}}
\newcommand{\nn}{\nonumber}

\newcommand{\rv}{\rho_{\rm vac}}
\newcommand{\rve}{\rho_{\rm vac}^{\rm eff}}
\newcommand{\ke}{\kappa_{\rm eff}}
\renewcommand{\mp}{m_P}
\newcommand{\mpe}{m_P^{\rm eff}}
\newcommand{\phiv}{\langle\phi\rangle}

\begin{document}

\baselineskip 14 pt
\parskip 12 pt

\begin{titlepage}

\begin{center}

\vspace{5mm}

{\Large \bf Pseudo-redundant vacuum energy}

\vspace{5mm}

Puneet Batra\footnote{pbatra@phys.columbia.edu}, Kurt Hinterbichler\footnote{kurth@phys.columbia.edu},
Lam Hui\footnote{lhui@astro.columbia.edu} and Daniel Kabat\footnote{kabat@phys.columbia.edu}

\vspace{2mm}

{\small \sl Institute for Strings, Cosmology and Astroparticle Physics} \\
{\small \sl and Department of Physics} \\
{\small \sl Columbia University, New York, NY 10027 USA}

\end{center}

\vskip 1.0 cm

\noindent
We discuss models that can account for today's dark energy.  The
underlying cosmological constant may be Planck scale but starts as a
redundant coupling which can be eliminated by a field redefinition.
The observed vacuum energy arises when the redundancy is explicitly
broken, say by a non-minimal coupling to curvature.  We give a recipe
for constructing models, including $R + 1/R$ type models, that realize
this mechanism and satisfy all solar system constraints on gravity.  A
similar model, based on Gauss-Bonnet gravity, provides a technically
natural explanation for dark energy and exhibits an interesting
see-saw behavior: a large underlying cosmological constant gives rise
to both low and high curvature solutions.  Such models could be
statistically favored in the string landscape.

\end{titlepage}
\setcounter{footnote}{0}

%%%%%%%%%%%%%%%%%%%%%%%%%%%%%%%%%%%%%%%%%%%%%%%%%%%%%%%%%%%%%%%%%%
\section{Introduction\label{Intro}}
%%%%%%%%%%%%%%%%%%%%%%%%%%%%%%%%%%%%%%%%%%%%%%%%%%%%%%%%%%%%%%%%%%

The present-day universe is undergoing exponential expansion
on a timescale set by today's Hubble parameter $H_0 \sim 10^{-33} \,
{\rm eV}$.  One of the biggest challenges in physics is understanding
why $H_0$ is so small relative to the energy scales of particle physics.
This is the puzzle of the cosmological constant, reviewed in
\cite{Weinberg:1988cp,Weinberg:1996xe,Carroll:2000fy,Weinberg:2000yb,
Polchinski:2006gy}.

Many solutions have been proposed \cite{Nobbenhuis:2004wn}.  An
approach which is currently popular is to invoke landscape / anthropic
ideas: in a potential with enough minima, at least one should have the
observed vacuum energy \cite{Bousso:2000xa}.  Although this may
ultimately prove to be the right explanation, it's still important to
explore alternatives.  In fact, the existence of a landscape may require
us to explore alternatives.  To explain this point, let us note that
in a simple model of flux compactification one has the potential
\cite{Bousso:2000xa,Douglas:2006es}
\[
V \sim -V_0 + \sum_{i = 1}^N c_i n_i^2\,.
\]
Here $-V_0$ is a large fixed negative energy density, $N$ is the
number of non-trivial cycles on some compactification manifold, the
$c_i$ are order one constants, and the integers $n_i$ measure the
fluxes through the various cycles.  The number of flux vacua then
grows with the vacuum energy roughly as
\be\label{minima}
\hbox{\# vacua} \sim (V + V_0)^{N/2}\,.
\ee
For large $N$ the number of vacua grows extremely rapidly with energy.
Now suppose there is a mechanism for (nearly) cancelling a large
positive vacuum energy, leaving a small effective cosmological
constant.  We do not expect to find such a mechanism without
introducing small parameters \cite{Weinberg:1988cp}.  But in view of
the growth (\ref{minima}), such a mechanism could be statistically
favored in the landscape, even though it requires carefully adjusted
couplings.

This leads us to consider theories which nearly cancel a large
underlying vacuum energy.  The basic mechanism we shall explore is the
following.  Suppose there is a potential $V$ for a collection of
fields $\psi$.  We assume that $V(\psi)$ vanishes somewhere in field
space, but (as should be generic) we assume the derivatives of $V$ are
non-zero at that point:
\[
V \big\vert_{\psi_0} = 0 \quad {\rm but} \quad \left.{\partial V \over \partial \psi_i}\right\vert_{\psi_0} \not= 0 \quad
\hbox{\rm for some $i$.}
\]
Then in a neighborhood of $\psi_0$ we can redefine fields, introducing
a new field $\phi$ such that
\be
\label{IntroPotential}
V = \rv - \phi\,.
\ee
Here $\rv$ is a constant representing a large underlying vacuum
energy.  Clearly at the level of the potential (\ref{IntroPotential}),
$\rv$ has no meaning: it is a redundant coupling which can be
eliminated by a field redefinition
\be
\label{IntroPseudo}
\phi \rightarrow \phi + a \qquad\quad \rv \rightarrow \rv + a\,.
\ee
However generically this pseudo-symmetry will be explicitly broken by
other terms in the Lagrangian.  By choosing appropriate breaking
terms, we will be able to write down models that have de Sitter
solutions of arbitrarily small curvature, even in the presence of an
underlying Planck-scale vacuum energy.  The necessary breaking terms
will involve extremely small couplings.  In practice this allows us to
shuffle fine tunings away from the cosmological constant and into
other sectors of the theory.  Such tunings may be expected to arise in
the landscape, as we emphasized above.  Also by shifting small
parameters around, we will be able to construct models in which a
small cosmological constant is technically natural (stable under
radiative corrections).  Interestingly, these models also exhibit a
kind of see-saw mechanism, in that a large underlying cosmological
constant gives rise to multiple solutions, some with high curvature,
some with low curvature.

An outline of this paper is as follows.  In section
\ref{sect:redundant} we develop the idea of making vacuum energy
nearly redundant.  In section \ref{sect:F(R)} we present a
model-building recipe based on fields with non-minimal couplings to
the scalar curvature.  In section \ref{sect:Rphi2} we discuss a
particular example in more detail: a model with an $R\phi^2$ coupling
that can be thought of as a generalization of $R + 1/R$ gravity
\cite{Carroll:2003wy}.  In section \ref{sect:GB} we discuss another
example, based on Gauss-Bonnet gravity, which provides a technically
natural explanation for dark energy.  In the appendices we show that
all models we consider can be made compatible with cosmological and
solar system tests of gravity, and give further details on issues
of classical and radiative stability.  Readers who are interested in the
concrete examples could skip directly to sections \ref{sect:Rphi2} or
\ref{sect:GB}, which can be read independently of the rest of the paper.

Conventions: our metric signature is $(-+++)$.  Quantities which
appear in the underlying Lagrangian include the bare gravitational
coupling $\kappa = 1/\mp$ and the bare vacuum energy $\rv$.  Quantities
measured at low energies include the effective gravitational coupling
$\ke = 1/\mpe = (8 \pi G_N)^{1/2} \sim 10^{18} \, {\rm GeV}$ and the
effective vacuum energy $\rve \sim (10^{-3} \, {\rm eV})^4$.

%%%%%%%%%%%%%%%%%%%%%%%%%%%%%%%%%%%%%%%%%%%%%%%%%%%%%%%%%%%%%%%%%%%%%%%%%%%%%%%%%%%%%
\section{Redundant vacuum energy\label{sect:redundant}}
%%%%%%%%%%%%%%%%%%%%%%%%%%%%%%%%%%%%%%%%%%%%%%%%%%%%%%%%%%%%%%%%%%%%%%%%%%%%%%%%%%%%%

Conventionally one imagines that vacuum energy arises from an (effective)
potential $V(\psi)$ for a collection of matter fields $\psi_i$.  To find static solutions one
goes to a minimum of the potential $V'(\psi) = 0$ and finds the vacuum energy
\[
\rv = V(\psi) \big\vert_{V' = 0}\,.
\]
Note that $\rv$ is invariant under field redefinitions of the
$\psi_i$.  The effective action is
\[
S = \int d^4x \, \sqrt{-g} \left({1 \over 2 \kappa^2} R - \rv\right)
\]
and one has the usual tuning problem for vacuum energy: $\rv / \mp^4 \ll 1$.

As an alternative to the conventional scenario imagine a potential which crosses
through zero somewhere in field space.  That is, suppose for some $\psi_0$
\[
V(\psi_0) = 0 \quad {\rm but} \quad V'(\psi_0) \not= 0\,.
\]
We can redefine fields in a neighborhood of $\psi_0$ to set $V(\phi) = \rv - J \phi$.
(The potential may have critical points elsewhere in field space, in which case this redefinition
can only be made locally.)  For simplicity we assume a canonical kinetic term for $\phi$ --
an assumption we will relax in the next section -- and consider the action
\be
\label{redundant}
S = \int d^4x \, \sqrt{-g} \left({1 \over 2 \kappa^2} R
- {1 \over 2} \partial_\mu \phi \partial^\mu \phi + J \phi - \rv\right)\,.
\ee
The interesting thing about this action is that it's invariant under
\[
\phi \rightarrow \phi + a \qquad\quad \rv \rightarrow \rv + J a\,.
\]
This means the bare vacuum energy $\rv$ has no observable
consequences: it's a redundant coupling which can be eliminated by a
field redefinition \cite{Weinberg:1995mt}.  This is unlike the source $J$ which could
be absorbed by rescaling the field but would then reappear in the kinetic term.
Of course (\ref{redundant}) has no static classical solutions, rather there are only
cosmological solutions in which the scalar field rolls down the hill.
$\rv$ can be absorbed into initial conditions for the field -- say the
value of the field at which $\partial_0 \phi$ vanishes.

To obtain de Sitter solutions one could imagine deforming
(\ref{redundant}) by turning on other couplings.  For example one
could consider the theory
\be
\label{redundant2}
S = \int d^4x \, \sqrt{-g} \left(f(\phi) R
- {1 \over 2} \partial_\mu \phi \partial^\mu \phi - V(\phi)\right)\,.
\ee
Now we can find de Sitter solutions.  Indeed for constant fields the scalar
equation of motion requires
\be
\label{redundant3}
{df \over d\phi} R = {dV \over d\phi} \qquad \hbox{\rm i.e.} \qquad R = {dV \over df}
\ee
and the Friedmann equation is
\be
\label{redundant4}
R = 2 V / f\,.
\ee
Provided $dV/d\phi$ and $df/d\phi$ are non-zero these equations have
novel de Sitter solutions: novel in the sense that (unlike the conventional
scenario) the effective vacuum energy is not determined by critical points
of the potential, and (unlike quintessence) our models do not invoke a
slowly-rolling scalar field.  We will refer to vacuum energy in our models
as being ``pseudo-redundant.''

What does making vacuum energy pseudo-redundant achieve?  We should state
at the outset that it does not solve the fine-tuning problems associated
with the cosmological constant.  As we will discuss in more detail in the
next section, the model (\ref{redundant2}) will have an effective Planck mass
$\mpe$ and an effective vacuum energy $\rve$, and we require
\[
\epsilon = {3 H^2 \over (\mpe)^2} = {\rve \over (\mpe)^4} \ll 1
\]
for the model to be compatible with observation.  This can be
achieved, but only by tuning parameters, in accord with Weinberg's
no-go theorem \cite{Weinberg:1988cp}.  However, as is manifest from
the construction, by making vacuum energy pseudo-redundant we have
shifted the necessary tunings away from the underlying cosmological
constant and into other sectors of the theory.  This freedom may be
important in the landscape.  Also, as we will see in section
\ref{sect:GB}, it will allow us to build technically natural models
for vacuum energy.\footnote{Full disclosure: models such as
(\ref{redundant2}) can be put in Einstein frame via a conformal
transformation, see appendix \ref{F(R):stability}, and in Einstein
frame the necessary tuning has the conventional
cosmological-constant form.  This is further motivation for
considering models which do not have an Einstein frame description,
such as the Gauss-Bonnet model of section \ref{sect:GB}.}

%%%%%%%%%%%%%%%%%%%%%%%%%%%%%%%%%%%%%%%%%%%%%%%%%%%%%%%%%%%%%%%%%%%%%%%%%%%%
\section{An $F(R)$ model-building recipe\label{sect:F(R)}}
%%%%%%%%%%%%%%%%%%%%%%%%%%%%%%%%%%%%%%%%%%%%%%%%%%%%%%%%%%%%%%%%%%%%%%%%%%%%

In this section we consider models which generalize (\ref{redundant2}) by including
a non-minimal kinetic term for the scalar,
\be
\label{F(R):action}
S = \int d^4x \, \sqrt{-g} \left(f(\phi) R
- {1 \over 2} h(\phi) \partial_\mu \phi \partial^\mu \phi - V(\phi)\right) + S_{\rm matter}
\ee
where matter is minimally-coupled to $g_{\mu\nu}$.  Models of this type, with applications to dark energy, have
been reviewed in \cite{Nojiri:2006ri,Woodard:2006nt}.  We will refer to them as $F(R)$ models.\footnote{In the literature
$F(R)$ gravity is usually taken to mean no scalar kinetic term, i.e.\ $h = 0$ in (\ref{F(R):action}).  As we
discuss below, our models agree with $F(R)$ gravity on cosmological solutions but differ inside the
solar system.}

As in the previous section, de Sitter solutions obey
\be
\label{F(R):solution}
R = {dV \over df} \qquad {\rm and} \qquad R = 2 V / f\,.
\ee
>From the action we can read off the effective Planck mass
\be
\label{F(R):mpe}
(\mpe)^2 = 2 \langle f(\phi) \rangle
\ee
and the effective vacuum energy
\be
\label{F(R):rve}
\rve = \langle V(\phi) \rangle\,.
\ee
We want to choose the functions $f$ and $V$ so that these quantities
take on their observed values.  However we also want to construct
models in which the de Sitter solutions are (meta-) stable and
compatible with solar system tests of gravity.  One way to achieve
this is to make $h$ large.  This does not affect the de Sitter
solutions, but it suppresses any spatial or temporal variation of
$\phi$, so for sufficiently large $h$ the model passes all solar and
stability tests and does not significantly alter the cosmological history during matter/radiation domination.  Stability is studied in more detail in appendix
\ref{F(R):stability}, solar tests are analyzed in appendix
\ref{F(R):solar}, and cosmology during matter/radiation domination is studied in appendix \ref{Rphi2:cosmology}.

>From now on we will assume $h$ is large enough to pass solar system, cosmology,
and stability tests, and we turn our attention to constructing models
that yield the desired low curvature de Sitter expansion.  We could
analyze this using the scalar-tensor form (\ref{F(R):action}).  But it
is convenient to assume constant $\phi$ and integrate out the scalar
field, to obtain the effective action for gravity
\be
\label{redundant5}
S = \int d^4x \sqrt{-g} \, {\cal L}(R) \qquad\quad {\cal L}(R) = f R - V\,.
\ee
Here $f$ and $V$ are implicitly functions of the curvature scalar,
determined by solving $R = dV / df$ to find $f = f(R)$.  This is
simply a Legendre transform.\footnote{Integrating out the scalar
is analogous to starting from a first-order Lagrangian
$p\dot{q} - H$ and integrating out the momentum using the equation
of motion $\dot{q} = {\partial H \over \partial p}$.}  The transformation
can be inverted with \cite{Nojiri:2003ft,Chiba:2003ir}
\be
\label{redundant6}
f = {d{\cal L} / dR} \qquad\quad
V = f R - {\cal L}
\ee
to go back to the scalar-tensor description.  As usual $V$ here can be
expressed purely in terms of $f$ using the relation $f = {d{\cal L}/dR}$.
Strictly speaking this gives us the scalar-tensor action
in terms of $R$ and $f$; we can then imagine an arbitrary functional
relationship between $f$ and $\phi$.

It's convenient to write the effective gravity action in terms of
\be
y = R / 4 (\mpe)^2
\ee
and set\footnote{Here $\mpe$ is the fixed constant defined in (\ref{F(R):mpe}).}
\be
{\cal L}(R) = (\mpe)^4 F(y)\,.
\ee
What conditions do we need to impose on $F(y)$ to obtain a
low-curvature de Sitter solution?  The Friedmann equation
(\ref{redundant4}) fixes $R = 2 V / f$, so we must have
\[
(\mpe)^4 F(y) = f R - V = V.
\]
This equation should hold at $y = \epsilon$, where $\epsilon = R_0 / 4
(\mpe)^2$, $R_0$ being the curvature today.  Identifying $\langle V
\rangle = \rve = \epsilon (\mpe)^4$ we have the condition $F(\epsilon)
= \epsilon$.  This will give the right radius of curvature in Planck
units.  But to get the right effective Planck mass we also need
$\langle f \rangle = {1 \over 2} (\mpe)^2$ which from the first
equation in (\ref{redundant6}) means that $F'(\epsilon) = 2$.

To summarize, any function $F(y)$ with
\be
\label{Fcond}
F(\epsilon) = \epsilon \qquad\quad
F'(\epsilon) = 2
\ee
will produce a low-curvature de Sitter solution with the right
effective Planck mass.  Clearly there are an infinite number of
functions which satisfy (\ref{Fcond}).  We present a number of
examples in appendix \ref{appendix:examples}, and study a particular
example -- the $R\phi^2$ model -- in more detail in the next section.
Some of these examples might come as a surprise. For instance,
models such as ${\cal L} \sim (1/R) + (1/R)^2$, where
an Einstein Hilbert term is completely absent, are actually viable
both in the cosmological and the solar system contexts.
Interested readers are urged to consult the appendix.

Here we make some general comments on tuning issues.  Any function
satisfying (\ref{Fcond}) is fine-tuned, in that its value near zero is
so small.  This criticism applies, for example, to ordinary gravity
with a cosmological constant, which corresponds to $F(y) = 2y -
\epsilon$.  This particular solution to (\ref{Fcond}) does not
interest us so much, because we are more interested in models in which
$F(y)$ has an ${\cal O}(1)$ constant piece representing a Planck-scale
bare vacuum energy.
\be
\label{rvcond}
F(y) \sim - 1 + \cdots
\ee
Satisfying both (\ref{Fcond}) and (\ref{rvcond}) requires an even
greater degree of fine-tuning than usual, since (\ref{Fcond}) by
itself suggests that $F(0) \approx - \epsilon$, while (\ref{rvcond})
by itself suggests that $F(0) \approx -1$.\footnote{As in appendix
\ref{appendix:examples} one could consider models with inverse powers
of $y$.  However this does not help with tuning as in such models $F(0)$
diverges.}  That is, in $F(R)$ gravity it is unnatural to have a small
effective vacuum energy.  But it is doubly unnatural to obtain a small
effective vacuum energy by nearly cancelling off a large underlying
vacuum energy: to achieve this requires an even greater degree
of fine-tuning than is usually associated with the cosmological
constant.  Nonetheless we present some explicit examples of $F(R)$
gravity with Planck-scale bare vacuum energy in appendix
\ref{appendix:examples}, and we discuss one model in more detail in
the next section.  Readers who are concerned with tuning issues are
advised to skip to the Gauss-Bonnet model of section \ref{sect:GB}.

It is also worth commenting on the difference between our $F(R)$
models and the conventional $F(R)$ models often found in the
literature (e.g.\ \cite{Carroll:2003wy,Hu:2007nk} and references
therein).  The latter models set the bare vacuum energy to zero by
hand (for instance $\rv$ would be set to zero in the $R\phi^2$ example
we consider next).  Moreover these models have no kinetic term for the
scalar field in Jordan frame, i.e.\ $h = 0$ in (\ref{F(R):action}).
This implies that even for the same $F(R)$, our predictions for
subhorizon fluctuations would be quite different from the conventional
model, even though we agree on the cosmological background.  In
general, our predictions are much closer to those of general
relativity, although there could be detectable deviations in solar
system tests.

Finally, let us point out that the model-building recipe we have
presented can be easily extended.  For example, one natural extension
is to demand that the solution be stable.  In terms of the
Einstein-frame potential $\widetilde{V} = {V \over (2 (\ke)^2 f)^2}$
defined in (\ref{stability2}), stability can be achieved by having
$\widetilde{V}'' > 0$.  Evaluated on the solution
(\ref{F(R):solution}), $\widetilde{V}'' > 0$ is equivalent to $V'' > 8
\epsilon$.  Noting that $V'' = 1/\left(d^2 {\cal L} / d R^2\right)$,
the condition for stability is simply $0 < F''(\epsilon) < 2 /
\epsilon$.  More generally, a program for reconstructing $F(R)$ gravity
from a given expansion history has been developed \cite{Woodard:2006nt,Boisseau:2000pr}.

%%%%%%%%%%%%%%%%%%%%%%%%%%%%%%%%%%%%%%%%%%%%%%%%%%%%%%%%%%%%%%%%%%
\section{$R \phi^2$ model\label{sect:Rphi2}}
%%%%%%%%%%%%%%%%%%%%%%%%%%%%%%%%%%%%%%%%%%%%%%%%%%%%%%%%%%%%%%%%%%

In this section we consider a particular example of $F(R)$ theory, the
$R\phi^2$ model with action
\be
\label{Rphi2}
S = \int d^4x \, \sqrt{-g} \left({1 \over 2 \kappa^2} R
- {1 \over 2} \partial_\mu \phi \partial^\mu \phi - {1 \over 2} \xi R \phi^2 + J \phi
 - \rv\right)\,.
\ee
We derive this action from our general recipe in appendix
\ref{appendix:examples}, but with the couplings $\kappa$, $\xi$, $\rv$
already adjusted so as to give the correct de Sitter radius.  Here,
rather than use our previous results, we will analyze the model from
scratch.  Note that we have set $h$, the kinetic coefficient in
(\ref{F(R):action}), to unity. A large $h$ is equivalent to a small
$\xi$ through a field redefinition.

The action has a term linear in $\phi$ which involves a source $J$.
We will not specify the dynamics that give rise to $J$, aside from
noting that it could arise from a Yukawa coupling $\bar{\psi} \psi
\phi$ with a condensate $\langle\bar{\psi}\psi\rangle \not= 0$.  We
have included a non-minimal coupling to curvature ${1 \over 2} \xi R
\phi^2$.  Also we have tuned the underlying $\phi$ mass to zero; as
can be seen below, we require $m_\phi^2 < \xi H^2$ for the model to
work.  More generally any additional potential for $\phi$, if present,
would have to be extremely flat.  Finally, particle physics could be
taken into account by minimally-coupling the standard model to
$g_{\mu\nu}$; any particle physics contributions to the vacuum energy
can be lumped into $\rv$.

To find cosmological solutions it's convenient to solve for $\phi$ in
terms of the source, $\phi = J/\xi R$ where we have assumed that $\phi$
is constant.  Plugging this back into the action we get \be
\label{R1overR}
S = \int d^4x \, \sqrt{-g} \left({1 \over 2 \kappa^2} R
+ {J^2 \over 2 \xi R}  - \rv \right)\,.
\ee
Aside from the vacuum energy term, this is the generalized gravity
model of Carroll et.~al.~\cite{Carroll:2003wy}.  The Friedmann
equation for this model is
\[
H^2 = {1 \over 3} \kappa^2 \rv - {\kappa^2 J^2 \over 48 \xi H^2}
\]
where we have assumed a constant scalar curvature $R = 12 H^2$, appropriate
for de Sitter space in a flat FRW slicing.  This quadratic equation for $H^2$
has solutions
\be
\label{quadratic}
H^2 = {\kappa^2 \over 6} \left[\rv \pm \sqrt{\rv^2 - {3 J^2 \over 4 \xi \kappa^2}}\right]\,.
\ee
Suppose $J^2 \ll \xi \kappa^2 \rv^2$.  Then there is the conventional or
high-curvature solution in which the Einstein-Hilbert term dominates
and $H^2 \approx {1 \over 3} \kappa^2 \rv$.  But there is also a
low-curvature solution in which the $1/R$ term dominates and $H^2
\approx J^2 / 16 \xi \rv$.  For $\xi \rv > 0$ this provides a
mechanism by which a large underlying vacuum energy can drive a slow
Hubble expansion.  We will refer to such pairs of high- and
low-curvature solutions as see-saw solutions for dark
energy.\footnote{Similar low-curvature solutions, driven by a large
matter energy density, were found in \cite{Hu:2007nk}.  In our
general model-building recipe of section \ref{sect:F(R)} there is no
real need to demand that solutions come in see-saw pairs, although
such pairs could arise.}

It is important to note that to obtain the effective $R + 1/R$
description of the $R\phi^2$ model we had to integrate out a light
scalar with a mass set by $R$.  So the action (\ref{R1overR}) provides
an effective description of cosmology but cannot be used on sub-Hubble
distances, where one must revert to the underlying scalar-tensor
theory (\ref{Rphi2}).  Indeed by plugging a constant vev for $\phi$
into the action (\ref{Rphi2}) we can read off the effective Planck
mass
\be
\label{EffectivePlanck}
\left(\mpe\right)^2 = {1 \over \kappa^2} - \xi \langle \phi \rangle^2
\ee
and the effective vacuum energy
\be
\label{EffectiveVac}
\rve = \rv - J \langle \phi \rangle\,.
\ee
Assuming constant $\phi$, the Friedmann equation is
\be
\label{Rphi2eom}
\left({1 \over \kappa^2} - \xi \phi^2\right) H^2 = {1 \over 3} \left(\rv - J \phi\right)
\ee
and the $\phi$ equation of motion fixes $\phi = J / 12 \xi H^2$.
Note that (\ref{Rphi2eom}) can be rewritten as
\[
\left(\mpe\right)^2 H^2 = {1 \over 3} \rve
\]
which is the usual Friedmann equation in the presence of vacuum
energy.  So in terms of the underlying $R\phi^2$ theory, given $\mpe
\sim 10^{18} \, {\rm GeV}$ and $H \sim 10^{-33} \, {\rm eV}$, the
scalar field will adjust its vev to cancel the bare $\rv$ and give
$\rve \sim \left(10^{-3} \, {\rm eV}\right)^4$.

The model has four adjustable couplings ($\kappa$, $J$, $\xi$ and $\rv$)
and accounts for two observed quantities ($\ke$ and $H$).  So we can
express two of our couplings -- say $\kappa$ and $J$ -- in terms of $\xi$,
$\rv$ and the measured parameters $\ke$, $H$ and $\rve \equiv 3
\left(\mpe\right)^2 H^2$.
\bea
\label{mpeqn}
&& \left(\mp\right)^2 = {1 / \kappa^2} = {1 \over 12 H^2} \left(\rv + 3 \rve\right) \\
\label{Jeqn}
&& J^2 = 12 \xi H^2 \left(\rv - \rve\right)
\eea
The scalar vev can likewise be expressed as
\be
\label{phieqn}
\phi^2 = {1 \over 12 \xi H^2} \left(\rv - \rve\right)\,.
\ee
Note that (\ref{Jeqn}) requires $\xi \rv > 0$.\footnote{Although we
are most interested in having $\xi$ and $\rv$ positive, there are
consistent solutions in which $\xi$, $\rv$ and the bare $\kappa^2$ are
all negative.}  Just to plug in some numbers, if we take $\xi = 1$
and $\rv = \left(10^{18} \, {\rm GeV}\right)^4$ we would have
\be
\label{Rphi2values}
J \sim \left(10 \, {\rm MeV}\right)^3 \qquad \mp \approx \langle \phi \rangle \sim 10^{86} \, {\rm eV}\,.
\ee

A few comments on these results:
\begin{enumerate}
\item
The see-saw mechanism requires $J^2 \ll \xi \kappa^2 \rv^2$.  Here
however we are interested in the regime $\rv \gg \rve$.  In this
case (\ref{mpeqn}) and (\ref{Jeqn}) imply $J^2 \approx \xi \kappa^2
\rv^2$, so in fact the see-saw mechanism isn't operative.  Rather we
have an approximate cancellation of the vacuum energy, $\rve \ll
\rv$, but also an approximate cancellation of the Planck mass, $\mpe
\ll \mp$, such that curvatures of the two solutions in
(\ref{quadratic}) are roughly the same.  Indeed for $J^2 \approx \xi
\kappa^2 \rv^2$ (\ref{quadratic}) reduces to
\[
H^2 \approx {1 \over 6} \kappa^2 \rv \left(1 \pm {1 \over 2}\right)\,.
\]
The see-saw regime of the $R\phi^2$ model is actually rather unphysical,
as one can show that the low-curvature see-saw solution has both $\rve < 0$
and $(\mpe)^2 < 0$.\footnote{We are grateful to Ali Masoumi for pointing
out this pathology.}
\item
The model can be made compatible with solar system tests of gravity.
This is unlike the original $R + 1/R$ models of cosmology which, if
naively extended to the solar system, have been ruled out
\cite{Chiba:2003ir,Chiba:2006jp}.  As we discuss in more detail in
appendix \ref{F(R):solar}, the $R\phi^2$ model can be related to
Brans-Dicke theory, and solar tests require
\be
\label{Rphi2omega}
\omega = {\rve \over \xi \left(\rv - \rve\right)} \gtrsim 10^5\,.
\ee
With a natural value for $\rv$ this means $\xi$ must be extremely small,
perhaps $\vert \xi \vert \lesssim 10^{-125}$; the sign of $\xi$ is correlated
with the sign of $\rv$.  Then the vev of $\phi$ becomes much larger than the
Planck mass, $\phi \sim \mp / \sqrt{\xi}$, which raises an important
issue: our conclusions are sensitive to the presence of additional
non-renormalizable operators suppressed by powers of $\phi/\mp$.
\item
As we show in appendix \ref{F(R):stability}, the low-curvature de
Sitter solution can be made metastable over cosmological timescales.
Interestingly, the resulting requirement on $\xi$ is similar to
that from solar system constraints. Therefore, according to this model,
if the dark energy equation of state is observed to differ from $-1$,
one should also expect to see general relativity violations just around
the corner in solar system tests.
\item
Achieving $\rve \ll \rv$ and $\rve \ll (\mpe)^4$ requires severe
fine-tuning.  In particular (\ref{mpeqn}) can be rewritten as
\be
\label{mptuning}
{\rv \over \mp^4} \approx 16 \cdot {\rve \over \rv} \cdot {\rve \over (\mpe)^4} \sim 10^{-240}
\ee
where we've assumed $\rv \sim (\mpe)^4$.  Likewise (\ref{mpeqn}) and
(\ref{Jeqn}) can be combined to give
\be
\label{Jtuning}
{J^2 \mp^2 \over \xi \rv^2} \approx 1 + 2 {\rve \over \rv} = 1 + {\cal O}(10^{-120})\,.
\ee
These two conditions mean the $R\phi^2$ model is actually much more
fine-tuned than a pure cosmological constant.  This is in accord with
our general discussion of tuning in $F(R)$ models at the end of section \ref{sect:F(R)}.
\item As we show in appendix \ref{Rphi2:radiative}, the necessary tunings 
are destabilized by radiative corrections.  The easiest way to see this is
to map the model to Einstein frame, where no symmetry protects the vacuum energy.
\item
$\mpe$ sets the scale at which gravity becomes strongly coupled, so
one might argue that the natural scale for vacuum energy is
$\left(\mpe\right)^4$.  This is the value we adopted in
(\ref{Rphi2values}).  It leads to an amusing coincidence: with $\rv
\sim \left(\mpe\right)^4$ the total energy inside our Hubble volume,
estimated as $\rve / H^3$, is always comparable to the bare Planck
mass $\mp$.  One could very well criticize this choice, on the
grounds that $\rv$ can't know about $\phi$-dependent quantities such
as $\mpe$; this is related to the question of why the potential for
$\phi$ is so flat.  To address this one could imagine tying $\rv$ to
a non-gravitational scale such as the scale for supersymmetry
breaking.
\end{enumerate}

%%%%%%%%%%%%%%%%%%%%%%%%%%%%%%%%%%%%%%%%%%%%%%%%%%%%%%%%%%%%%%%%%%
\section{Gauss-Bonnet model\label{sect:GB}}
%%%%%%%%%%%%%%%%%%%%%%%%%%%%%%%%%%%%%%%%%%%%%%%%%%%%%%%%%%%%%%%%%%

The mechanism we have discussed is clearly rather general: one could
start with a model in which vacuum energy is redundant and explicitly
break the redundancy in any number of ways.  One possibility is to
introduce a scalar field with a non-minimal coupling to a more general
curvature invariant.  Upon integrating out the scalar, such a model
would yield one of the generalized modified gravity theories studied
in \cite{Carroll:2004de}.

A particularly appealing possibility is to couple the scalar field to
the Euler density $\mathcal{G}$.  Thus we consider the action
\be
\label{GBmodel}
S = \int d^4x \, \sqrt{-g} \left({1 \over 2 \kappa^2} R
- {1 \over 2} \partial_\mu \phi \partial^\mu \phi - {1 \over 2} \xi \kappa^2 \mathcal{G} \phi^2 + J \phi
- \rv \right)
\ee
where
\[
\mathcal{G} \equiv R^2 - 4 R_{\mu\nu} R^{\mu\nu}
+ R_{\mu\nu\lambda\sigma} R^{\mu\nu\lambda\sigma}\,.
\]
This is known as Gauss-Bonnet gravity.  For a recent review see
\cite{Neupane:2007qw}.  Gauss-Bonnet gravity has a number of desirable
features.  It gives a ghost-free theory with second-order equations of
motion \cite{Stelle:1977ry}.  It can be generated by $\alpha'$
corrections in string theory \cite{Zwiebach:1985uq}.  And most
importantly from our point of view, it will allow us to avoid many of
the fine-tuning problems of the $R \phi^2$ model.  One could imagine
adding a mass term for the scalar, but the mass would have
to be tiny: as can be seen below, we need $m^2 < \xi \kappa^2 H^4$ for
the model to work.

Assuming $\phi$ is constant we can integrate it out, $\phi = J / \xi
\kappa^2 {\mathcal G}$ to obtain the effective action
\[
S = \int d^4x \, \sqrt{-g} \left({1 \over 2 \kappa^2} R + {J^2 \over 2 \xi \kappa^2 {\mathcal G}} - \rv \right)\,.
\]
This is a particular example of the generalized modified gravity of
Carroll et.~al.\ \cite{Carroll:2004de}, supplemented with a vacuum
energy term.  On maximally-symmetric spaces ${\mathcal G} = R^2/6$.
So not surprisingly there are solutions in which the Gauss-Bonnet term
dominates over Einstein-Hilbert and $R^2 \sim J^2 / \xi \kappa^2 \rv$.

To analyze the model more carefully we should vary (\ref{GBmodel}) with
respect to $\phi$ and the metric then restrict to constant $\phi$ and
maximally symmetric metrics.  We do this in appendix \ref{appendix:GB}.
However if one is only interested in de Sitter solutions a general analysis
isn't necessary.  Recall that $\mathcal{G}$ is a total derivative in four
dimensions, so upon plugging a constant vev for $\phi$ into the action
(\ref{GBmodel}) the Gauss-Bonnet term can be dropped.  The effective Planck
mass is therefore unshifted from its bare value, $\mpe = \mp$, while the
effective vacuum energy is $\rve = \rv - J \langle \phi \rangle$.
Thus the Friedmann equation is
\be
\label{GBFriedmann}
R = 12 H^2 = 4 \kappa^2 \rve = 4 \kappa^2 (\rv - J \phi)
\ee
where the scalar equation of motion fixes
\be
\label{GBphieqn}
\phi = 6 J / \xi \kappa^2 R^2\,.
\ee
Eliminating $\phi$ gives a cubic equation for $R$.  The see-saw
mechanism operates when $J^2 \ll \xi\kappa^6\rv^3$.  To leading order in
this regime the three solutions are
\be
\label{GBseesaw}
R \approx 4 \kappa^2 \rv, \qquad\quad
R \approx \pm \sqrt{6 J^2 \over \xi \kappa^2 \rv}\,.
\ee
We require that $\xi$ and $\rv$ have the same sign so that all three
roots are real.  Then we have one solution with large curvature and
two solutions with small opposite-sign curvatures.

In the small-curvature de Sitter solution we can solve for $J$ in
terms of $\xi$, $\rv$ and the observed quantities $\kappa$ and $H$.
\be
\label{GBJeqn}
J^2 = 24 \xi \kappa^2 H^4 \rv
\ee
Likewise the vev is given in terms of these quantities by
\be
\label{GBvev}
\langle \phi^2 \rangle = \rv / 24 \xi \kappa^2 H^4\,.
\ee
Just to plug in some numbers, setting $\xi = 1$ and $\rv = (10^{18} \,
{\rm GeV})^4$ we have
\[
J \sim (10^{-13} \, {\rm eV})^3 \qquad\quad \langle \phi \rangle \sim 10^{146} \, {\rm eV}\,.
\]

A few comments on this model:
\begin{enumerate}
\item
Unlike the $R\phi^2$ model, the Gauss-Bonnet model realizes the
see-saw mechanism in the regime of interest: an underlying
Planck-scale vacuum energy can drive either a fast or a slow de
Sitter expansion.
\item
There is no need to fine-tune the bare Planck mass, nor is there
any need to take the bare $\rv \ll \mp^4$.  In this sense the
Gauss-Bonnet model avoids some of the fine-tunings necessary in the
$R\phi^2$ model.
\item
The only small number we need in Gauss-Bonnet is $J^2 \ll
\xi\kappa^6\rv^3$.  Note that (\ref{GBJeqn}) implies
\be
\label{GBtuning}
{J^2 \mp^6 \over \xi \rv^3} \sim \left(\rve \over \rv\right)^2 \sim 10^{-240}\,.
\ee
As we will show in section \ref{GB:radiative}, this small parameter is stable
under radiative corrections!
\item
The Gauss-Bonnet model does, however, involve super-Planckian
vevs.  This makes the model sensitive to any other higher-dimension
operators that might be present in the Lagrangian.
\end{enumerate}

%%%%%%%%%%%%%%%%%%%%%%%%%%%%%%%%%%%%%%%%%%%%%%%%%%%%%%%%%%%%%%%%%%%%%%%%%%%%%%%%%%%%%%%%
\subsection{Classical stability and solar system tests\label{GB:solar}}
%%%%%%%%%%%%%%%%%%%%%%%%%%%%%%%%%%%%%%%%%%%%%%%%%%%%%%%%%%%%%%%%%%%%%%%%%%%%%%%%%%%%%%%%

In appendix \ref{appendix:GB} we show that the low-curvature de Sitter solution is
metastable over cosmological timescales provided $\xi \lesssim 1$.  However we
also want the model to be compatible with solar system tests.  Without
performing a full analysis, we believe that for sufficiently small $\xi$
the model will pass all solar system tests of gravity.  The essential
point is that the scalar field can be made arbitrarily weakly coupled.
To see this take the Gauss-Bonnet action (\ref{GBmodel}) and set
\be
\phi = \phiv + \delta \phi \qquad \quad g_{\mu\nu} = \langle g_{\mu\nu} \rangle + \delta g_{\mu\nu}\,.
\ee
Dropping a total derivative, the full Gauss-Bonnet action becomes
(here $\rve$ is a fixed numerical quantity, and $R$ for example means
$\langle R \rangle + \delta R$)
\be
\label{shiftedGB}
S = \int d^4x \sqrt{-g} \left({1 \over 2 \kappa^2} R - \rve - {1 \over 2} \partial_\mu (\delta\phi)
\partial^\mu (\delta\phi) - \xi \kappa^2 \phiv \, \delta {\mathcal G} \, \delta \phi
- {1 \over 2} \xi \kappa^2 {\mathcal G} (\delta \phi)^2\right)
\ee
The last two terms represent non-minimal couplings between $\delta
\phi$ and the metric.  But for fixed $\rv$, (\ref{GBvev}) implies that
$\phiv \sim 1/\sqrt{\xi}$.  So as $\xi \rightarrow 0$ these
non-minimal couplings vanish (they scale as $\xi^{1/2}$ and $\xi$,
respectively).  Thus as $\xi \rightarrow 0$ the model goes over to
Einstein gravity with a cosmological constant plus a massless,
minimally-coupled scalar.  Such a model is compatible with all solar
system tests of gravity, so we expect Gauss-Bonnet to pass solar tests
provided $\xi$ is sufficiently small.  This does not address the
question of exactly how small $\xi$ must be; to settle this question
would require a more detailed analysis along the lines of
\cite{Amendola:2007ni}.\footnote{The analysis in
\cite{Amendola:2007ni} has a restricted range of validity, see their
(20), and cannot be directly applied to our model.}

%%%%%%%%%%%%%%%%%%%%%%%%%%%%%%%%%%%%%%%%%%%%%%%%%%%%%%%%%%%%%%%%%%%%%%%%%%%%%%%%%%%%%%%%
\subsection{Radiative corrections\label{GB:radiative}}
%%%%%%%%%%%%%%%%%%%%%%%%%%%%%%%%%%%%%%%%%%%%%%%%%%%%%%%%%%%%%%%%%%%%%%%%%%%%%%%%%%%%%%%%

Finally we consider radiative corrections in the Gauss-Bonnet model.
We will argue that radiative corrections to the scalar potential are
under control provided $\xi$ is sufficiently small, roughly $\xi
\lesssim 10^{-240}$.  In this sense, for small $\xi$, the Gauss-Bonnet
model provides a technically natural explanation for dark energy.

When $\xi = J = 0$ the scalar field has a shift symmetry $\phi
\rightarrow \phi + {\rm const.}$ which forbids any corrections to the
scalar potential.  So one might expect radiative corrections to the
scalar mass, for example, to be proportional to $\xi$.  On dimensional
grounds one might expect a correction $\delta m^2 \sim \xi
\Lambda_{UV}^2$ to be generated.  If the UV cutoff scale $\Lambda_{UV}
\sim \mp$ this would be bad news, because the Gauss-Bonnet model
requires $m^2 < \xi \kappa^2 H^4$.

Here the remarkable structure of Gauss-Bonnet gravity comes to the
rescue.  Expanding about a de Sitter solution, the relevant
interaction vertices can be read off from the last two terms in
(\ref{shiftedGB}).  Since the Euler density is a total derivative,
{\em the vertices vanish when all scalar lines together carry no net momentum
into any vertex.}\footnote{Additional interaction vertices arise
from the scalar kinetic term, but due to the shift symmetry they
vanish at zero scalar momentum and we can ignore them.}  This means
graphs like (scalar lines are solid, graviton lines are dotted)
\begin{center}
{\includegraphics[height=2.5cm]{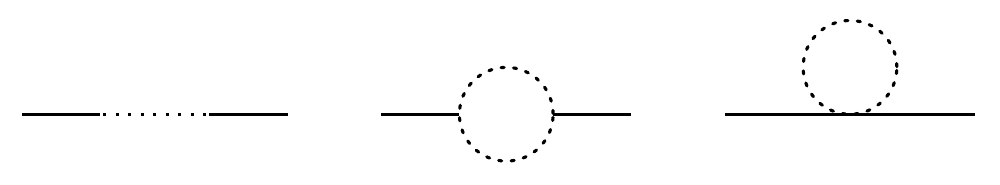}}
\end{center}
which naively generate an ${\cal O}(\xi)$ correction to the scalar
mass, actually vanish at zero external momentum.  The leading
correction to the scalar mass seems to come from a two-loop diagram.
\begin{center}
{\includegraphics[height=2cm]{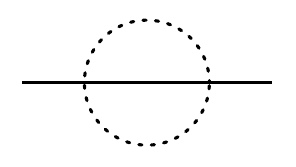}}
\end{center}
A simple estimate is that this diagram generates an ${\cal O}(\xi^2)$
correction\footnote{Each vertex has a coupling $\xi \kappa^2$ and
involves two curvatures, hence two powers of momentum for each
graviton.  Setting $g_{\mu\nu} = \langle g_{\mu\nu} \rangle + \delta
g_{\mu\nu}$ the graviton propagator $\sim\kappa^2 / p^2$.}
\bea
\delta m^2 & \sim & \xi^2 \kappa^4 \int{d^4p \over (2\pi)^4} {d^4q \over (2\pi)^4}\,
p^4 q^4 {\kappa^2 \over p^2} {\kappa^2 \over q^2} {1 \over (p+q)^2} \\
& \sim & \xi^2 \kappa^8 \Lambda_{UV}^{10}
\eea
With $\Lambda_{UV} \sim \mp$ this means $\delta m^2 \sim \xi^2 \mp^2$.
So radiative corrections to the scalar mass are under control for
\[
\xi^2 \mp^2 < \xi \kappa^2 H^4
\]
or equivalently
\[
\xi < (\rve / \mp^4)^2 \sim 10^{-240}\,.
\]

In a similar way one can study corrections to the linear source $J$.
Here preserving the relation (\ref{GBtuning}) requires that $\delta J
/ J < 1$.  As above, the leading correction to $J$ seems to come from
a two-loop tadpole.
\begin{center}
{\includegraphics[height=2cm]{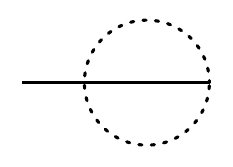}}
\end{center}
This generates a correction
\bea
\delta J & \sim & \xi \kappa^2 \xi \kappa^2 \phiv \int{d^4p \over (2\pi)^4} {d^4q \over (2\pi)^4}\,
p^4 q^4 {\kappa^2 \over p^2} {\kappa^2 \over q^2} {1 \over (p+q)^2} \\
& \sim & \xi^2 \kappa^8 \phiv \Lambda_{UV}^{10}
\eea
Taking $\Lambda_{UV} \sim \mp$, and using (\ref{GBphieqn}) to
eliminate the vev, this means
\[
{\delta J \over J} \sim {\xi \mp^4 \over H^4}\,.
\]
Again radiative corrections are under control for $\xi < 10^{-240}$.

Finally, what about radiative corrections to $\rv$ itself?  We expect radiative
corrections to generate a shift
\[
\delta \rv \sim \Lambda_{UV}^4 \sim \mp^4\,.
\]
Assuming we start with $\rv \sim (\mp)^4$, this means $\delta \rv \sim
\rv$.  This preserves the tuning (\ref{GBtuning}), so the model is
stable under radiative corrections to the vacuum energy.  More
directly, as can be seen from (\ref{GBFriedmann}) and
(\ref{GBseesaw}), the model will compensate for the change in $\rv$ by
shifting $\langle \phi \rangle$ so as to preserve a small effective
vacuum energy, with $\delta \rve \sim \rve$.  For this argument to
hold it's important that shifting $\langle \phi \rangle$ doesn't
change the Planck mass.  (In this respect the Gauss-Bonnet model
differs from the $R\phi^2$ model studied in appendix
\ref{Rphi2:radiative}.)

%%%%%%%%%%%%%%%%%%%%%%%%%%%%%%%%%%%%%%%%%%%%%%%%%%%%%%%%%%%%%%%%%%%%%%%%%%%%%%%%%%%
\section{Conclusions\label{Conclusions}}
%%%%%%%%%%%%%%%%%%%%%%%%%%%%%%%%%%%%%%%%%%%%%%%%%%%%%%%%%%%%%%%%%%%%%%%%%%%%%%%%%%%

In this paper we explored the idea that the underlying vacuum energy
could be large, of order the (effective) Planck scale, with the
observed slow Hubble expansion due to the dynamics of a scalar field
with non-minimal couplings.  We showed that the idea could be realized
in the $R\phi^2$ model and its $F(R)$ generalizations.  However all
such theories, involving only the scalar curvature, require severe
fine-tuning and are unstable with respect to radiative corrections.

We went on to consider the Gauss-Bonnet model in which the scalar
field couples to the Euler density.  The Gauss-Bonnet model requires
that one combination of parameters be small, namely $J^2 \ll
\xi\kappa^6\rv^3$.  However we argued that for small $\xi$ this parameter
is stable under radiative corrections, and in this sense the
Gauss-Bonnet model provides a technically natural explanation for dark
energy.

The crucial feature that makes all this possible is the fact that the
bare vacuum energy is a redundant coupling, and therefore
unobservable, when $\xi = 0$.  This allows us to shuffle the necessary
fine-tunings among various couplings in the Lagrangian.  The way in which
tunings are shuffled will play an important role in determining how likely
these models are in the landscape.  Also, by shifting some of the tuning
to the Gauss-Bonnet term, pseudo-redundancy made it possible to construct
radiatively stable models for dark energy.

This leaves many open questions:
\begin{itemize}
\item
Can one envisage a cosmological history in which the scalar
field naturally evolves to take on its required expectation value?
Or does a realistic cosmology require modifying the model in some
way?  If one does keep $\xi$ small throughout cosmological evolution
then the quantum diffusion studied in \cite{Linde,Garriga:2003hj}
will play an important role.
\item
Some models, in particular Gauss-Bonnet, realize a see-saw mechanism
in which a large underlying vacuum energy can drive either a fast or a
slow de Sitter expansion.  Can one find interesting cosmological solutions
in which the universe spends some time inflating near the high-curvature
solution before evolving to low curvature?
\item
The models we studied require super-Planckian vevs which are
often viewed as problematic \cite{McAllister:2007bg}.  But are these
large vevs really necessary?  Or could the models be modified in
some way to eliminate them?
\item
Can the necessary non-minimal couplings be realized in a
UV-complete theory such as string theory?  Or is there some
fundamental obstacle to achieving this?
\item
One intriguing feature of the $R\phi^2$ model is that cosmological
stability and solar system constraints put roughly similar
requirements on the coupling $\xi$. Therefore, if $\xi$ takes a
value such that the dark energy equation of state deviates from $-1$
by an observable amount, it is likely that one would also see
violations of general relativity in precision solar system
measurements. This offers an interesting way to test the model. It
would be useful to check if the same property holds in the
Gauss-Bonnet model.
\end{itemize}

\bigskip
\goodbreak
\centerline{\bf Acknowledgements}
\noindent
We are grateful to Matt Kleban, Ali Masoumi, Alberto Nicolis, Massimo Porrati and
Iggy Sawicki for valuable discussions. PB, KH, LH and DK are supported
by DOE grant DE-FG02-92ER40699 and by a Columbia University
Initiatives in Science and Engineering grant.

\appendix
%%%%%%%%%%%%%%%%%%%%%%%%%%%%%%%%%%%%%%%%%%%%%%%%%%%%%%%%%%%%%%%%%%%%%%%%%%%%%%%%%%%
\section{$F(R)$ dynamics\label{appendix:F(R)}}
%%%%%%%%%%%%%%%%%%%%%%%%%%%%%%%%%%%%%%%%%%%%%%%%%%%%%%%%%%%%%%%%%%%%%%%%%%%%%%%%%%%

In this appendix we study the dynamics of the $F(R)$ type models in more
detail.  We study classical stability of the de Sitter solutions,
discuss solar system and cosmological constraints, and analyze radiative corrections,
in some cases in general $F(R)$ gravity, and in some cases in the
context of the $R\phi^2$ model.

%%%%%%%%%%%%%%%%%%%%%%%%%%%%%%%%%%%%%%%%%%%%%%%%%%%%%%%%%%%%%%%%%%
\subsection{Classical stability\label{F(R):stability}}
%%%%%%%%%%%%%%%%%%%%%%%%%%%%%%%%%%%%%%%%%%%%%%%%%%%%%%%%%%%%%%%%%%

To study the stability of the $F(R)$ solutions it's useful to make a
conformal transformation.  For a generic scalar-tensor theory in
Jordan frame
\[
S = \int d^4x \sqrt{-g} \left\lbrace f(\phi) R - {1 \over 2} h(\phi) g^{\mu\nu}
\partial_\mu \phi \partial_\nu \phi - V(\phi) \right\rbrace
\]
we define $\tilde{g}_{\mu\nu} = 2 (\ke)^2 f(\phi) g_{\mu\nu}$, where $\ke$ is a
fixed numerical quantity, to go to the Einstein-frame action
\be
\label{stability1}
S = \int d^4x \sqrt{-\tilde{g}} \left\lbrace {1 \over 2 (\ke)^2} \tilde{R}
- {1 \over 2} \tilde{h}(\phi) \tilde{g}^{\mu\nu} \partial_\mu \phi
\partial_\nu \phi - \widetilde{V}(\phi) \right\rbrace\,.
\ee
In this frame the scalar field is minimally coupled and
\be
\label{stability2}
\tilde{h}(\phi) = {1 \over 2 (\ke)^2} \left({h \over f} + {3 (f^\prime)^2 \over f^2}\right) \qquad
\widetilde{V}(\phi) = {V(\phi) \over \big(2 (\ke)^2f\big)^2} \,.
\ee
By the definition of the effective Planck mass note that $f
\vert_{\langle \phi \rangle} = 1/2 (\ke)^2$.  Thus on our static
solution the Jordan-frame and Einstein-frame metrics are the same.  We
can then read off the dynamics of the scalar field just from its
kinetic term and potential.  The key point is that by making $h
\vert_{\langle \phi \rangle}$ large we can suppress any temporal
variation of the field and make the solutions (meta-) stable over an
arbitrarily long time scale.

For further discussion let's specialize to $R\phi^2$ model for
which\footnote{By a field redefinition, instead of making $h$ large,
we will set $h = 1$ and take $\xi$ to be small.}
\[
f(\phi) = {1 \over 2\kappa^2} \big(1 - \xi \kappa^2 \phi^2\big) \qquad
h(\phi) = 1 \qquad
V(\phi) = \rv - J \phi\,.
\]
In the regime of interest, where $\vert \xi \vert \ll 1$, we also have
$\tilde{h} \vert_{\langle \phi \rangle} \approx 1$.  So small
fluctuations about the static solution are governed by the effective
potential
\[
\widetilde{V} = \Big({\kappa \over \ke}\Big)^4 {\rv - J \phi \over (1 - \xi \kappa^2 \phi^2)^2}\,.
\]
The potential is sketched in Fig.~1.  It vanishes when $\phi = \rv /
J$ and blows up where the effective gravitational coupling diverges,
at $\phi^2 = 1 / \xi \kappa^2$.  The potential has two critical points,
with large and small vevs, corresponding to the two solutions in
(\ref{quadratic}).  We are interested in the large-vev solution.

\begin{figure}
\begin{center}
\raisebox{-3.44cm}{\includegraphics[height=7cm]{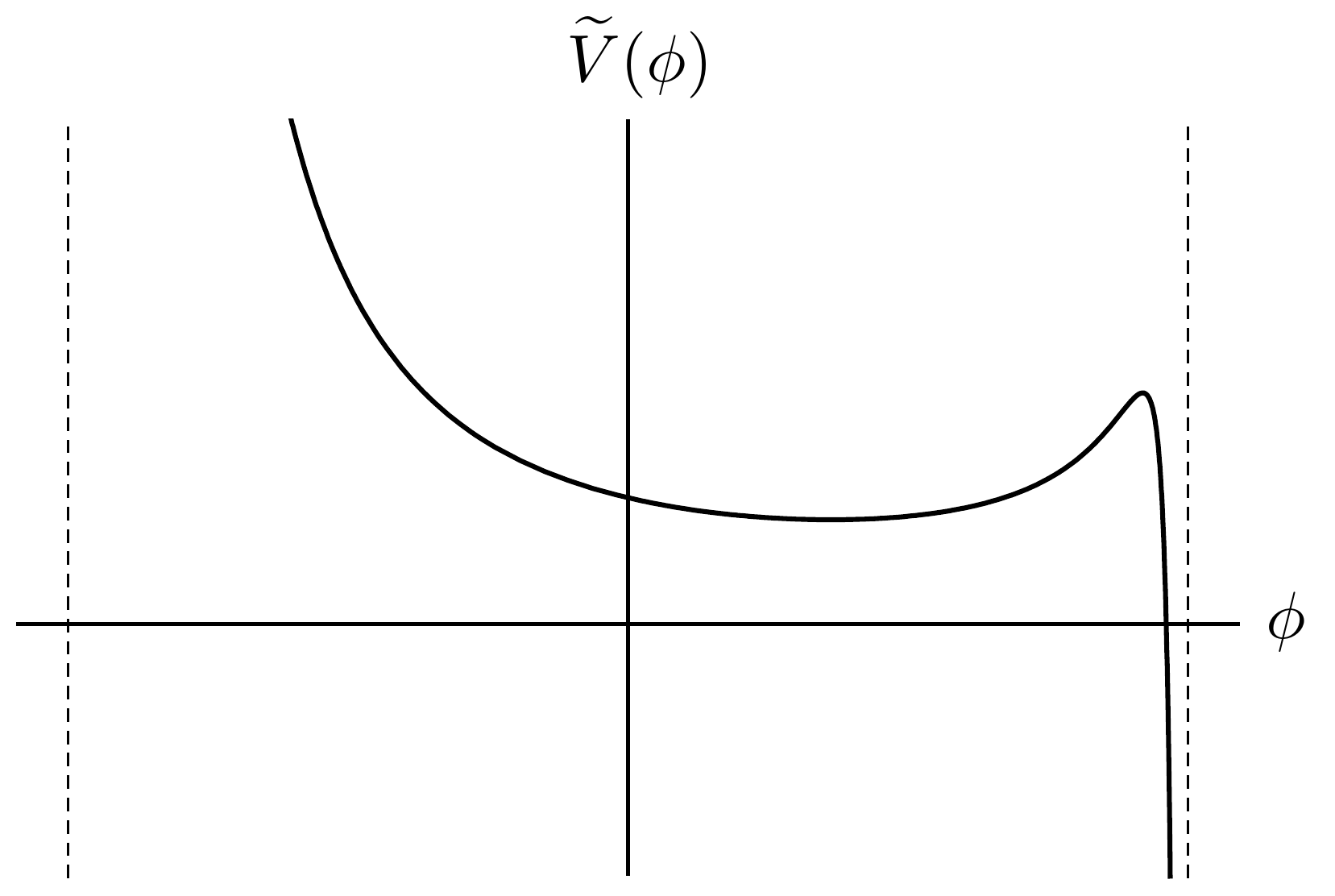}}
\end{center}
\caption{Effective potential for the $R\phi^2$ model.}
\end{figure}

To study stability we expand about the unstable point, $\phi = \langle
\phi \rangle + \delta \phi$.  To quadratic order the potential is (for
$\rve \ll \rv$)
\[
\widetilde{V}(\phi) = \rve - {\xi \rv \over (\mpe)^2} (\delta \phi)^2 + {\cal O}((\delta \phi)^3)\,.
\]
The timescale associated with the instability is
\[
\tau = \left(-\widetilde{V}^{\prime\prime}\right)^{-1/2} = \left({\rve \over 6 \xi \rv}\right)^{1/2} H^{-1}\,.
\]
So for sufficiently small $\xi$, say $\xi \lesssim 10^{-120}$, the
solution will be stable for more than a Hubble time (and Hubble
friction will play an important role in the evolution of $\phi$).

%%%%%%%%%%%%%%%%%%%%%%%%%%%%%%%%%%%%%%%%%%%%%%%%%%%%%%%%%%%%%%%%%%
\subsection{Solar system tests\label{F(R):solar}}
%%%%%%%%%%%%%%%%%%%%%%%%%%%%%%%%%%%%%%%%%%%%%%%%%%%%%%%%%%%%%%%%%%

In section \ref{sect:F(R)} we constructed $F(R)$ models that give the
right Planck length and de Sitter radius.  However we also want models
that are compatible with solar system tests of gravity.  One way to
achieve this is to give the scalar field a non-minimal kinetic term.
As in (\ref{F(R):action}) we take
\[
S = \int d^4x \, \sqrt{-g} \left(f(\phi) R
- {1 \over 2} h(\phi) \partial_\mu \phi \partial^\mu \phi - V(\phi)\right) + S_{\rm matter}\,.
\]
We assume that matter is minimally coupled to $g_{\mu\nu}$.  By taking
$h$ large any spatial variation in $\phi$ is suppressed and
the model can be made compatible with solar system tests of
gravity.  Note that by a field redefinition $\phi \rightarrow \phi /
\sqrt{\langle h \rangle}$ taking $h$ large is equivalent to making
the scalar field very weakly coupled, i.e.\ $\xi$ small in the
$R\phi^2$ context.

To show this in more detail we use the equivalence to Brans-Dicke
gravity \cite{Chiba:1999wt}.  The classic solar system constraints on
Brans-Dicke gravity were obtained by setting $V = 0$.  We can
therefore translate the known constraints to our situation as long as
terms related to $V$ in both the gravitational and scalar equations of
motion are subdominant in the solar system. The Einstein equations
read
\begin{eqnarray}
\label{F(R)Einstein}
&&G_{\mu\nu} = f^{-1} \left({1\over 2} T_{\mu\nu}^{\rm matter} + {1\over 2} T_{\mu\nu}^\phi
+ \nabla_{\mu} \nabla_{\nu} f - g_{\mu\nu} \Box f \right) \\ \nonumber
&&T_{\mu\nu}^\phi \equiv h \nabla_\mu \phi \nabla_\nu \phi - g_{\mu\nu}
\left( {1\over 2} h \nabla^\rho \phi \nabla_\rho \phi + V \right)
\end{eqnarray}
and the scalar equation of motion is
\be
h \Box \phi + {1\over 2} {dh\over d\phi} \nabla^\mu \phi \nabla_\mu \phi - {dV\over d\phi} + 
{df \over d\phi} R = 0\,.
\ee
>From these equations it seems we need
\be
\label{requirements}
fR \gg V \quad , \quad \left|{df\over d\phi}\right| R \gg \left|{dV \over d\phi}\right|
\ee
where $R$ is the curvature in the solar system.  Recall that our
static de Sitter solution has $f_0 R_0 = 2 V_0$ and $f'_0 R_0 = V'_0$,
where $R_0 \sim H_0^2 \ll R_{\rm solar}$, and $f_0$ and $V_0$ denote
cosmological vevs.  It therefore seems the above two inequalities are
easily satisfied.  However there is an implicit assumption here, that
the value of $\phi$ in the solar system does not differ too
significantly from its cosmological value, so that $f_{\rm solar}
\approx f_0$ and $V_{\rm solar} \approx V_0$.  This will be true for
sufficiently large $h$, as we now show.\footnote{As an alternative
approach to satisfying solar tests, imagine taking $h = 0$.  Then
the model is strictly equivalent to $F(R)$ gravity, even within the
solar system.  The scalar equation of motion $R = V'/f'$ requires
that the scalar field, or at least $V'/f'$, change significantly
inside the solar system.  Then with suitable $F(R)$ one can evade
solar tests of gravity via an analog of the chameleon mechanism
\cite{Hu:2007nk}.}

To study the variation of $\phi$ within the solar system we linearize
the scalar equation of motion, setting $\phi = \phiv + \delta \phi$
with $\delta \phi$ obeying
\be
\nabla^2 \delta \phi + {f_0'' R - V_0'' \over h_0} \delta \phi + {f_0'(R - R_0)
\over h_0} = 0\,.
\ee
For large $h_0$ we can neglect the mass term.  The source term is
dominated by the scalar curvature in the solar interior, where $R
\approx \ke^2 \rho_\odot$.  Thus to a good approximation
\be
\nabla^2 \delta \phi + {f_0' \ke^2 M_\odot \over h_0} \delta^3({\bf x}) = 0\,.
\ee
So $\delta \phi$ is maximized at the surface of the sun, where
\be
\delta\phi \approx {f_0' \ke^2 M_\odot \over 4 \pi h_0 r_\odot}\,.
\ee
>From (\ref{redundant3}) and (\ref{redundant4}) it follows that $\delta
V / V \approx 2 \, \delta \! f / f$, so we only need to impose
\be
\label{deltaVoverV}
{\delta V \over V} \approx {V_0' f_0' \ke^2 M_\odot \over 4 \pi h_0 V_0 r_\odot} \ll 1\,.
\ee
This is easily satisfied for large $h_0$.  Then the model is
equivalent to Brans-Dicke theory, and solar system tests mostly bound
two parameters \cite{Will:2005va}
\be
\label{WillSolar}
\gamma = {1 + \omega \over 2 + \omega} = 1 \pm O(10^{-5}) 
\quad , \quad \beta = 1 + {\omega' \over
(3 + 2\omega)^2 (4 + 2\omega)} = 1 \pm O(10^{-3})
\ee
where $\omega$ and $\omega'$ are related to our $f$ and $h$ by:
\begin{eqnarray}
&& \omega = {1\over 2} h f \left({df \over d\phi}\right)^{-2} \\
&& \omega' = 2f \left( {1\over 2} {dh\over d\phi} {f \over (df/d\phi)^3}
+ {1\over 2} {h \over (df/d\phi)^2} - {h f (d^2 f/d\phi^2) \over (df/d\phi)^4} \right)\,.
\end{eqnarray}
As long as $h$ is large enough the solar system tests are passed.

For example, as shown in \cite{Chiba:1999wt}, in the $R\phi^2$ model we have $h = 1$, $f = {1 \over 2
\kappa^2} - {1 \over 2} \xi \phi^2$, $V = \rv - J \phi$ and
therefore $\omega = {1 - \xi \kappa^2 \phi^2 \over 4 \xi^2 \kappa^2 \phi^2}$.  With the help of (\ref{phieqn}) this can be rewritten as
\be
\label{appendix:Rphi2omega}
\omega = {\rve \over \xi \left(\rv - \rve\right)}\,.
\ee
The bound (\ref{WillSolar}) requires $\omega \gtrsim 10^5$ (we take
$\omega > -3/2$ so that in Einstein frame the Brans-Dicke scalar has a
right-sign kinetic term).  This leads to the constraint on $\xi$ in
(\ref{Rphi2omega}).  However for this constraint to be valid we should
make sure that (\ref{deltaVoverV}) is satisfied.  Using $J\phi \approx
\rv$ (\ref{deltaVoverV}) can be rewritten as
\be
\xi \ll 4 \pi \cdot {\rve \over \rv} \cdot {r_\odot \over \ke^2 M_\odot} = 4 \pi \cdot 10^{-120} \cdot 10^5\,.
\ee
This is less stringent than (\ref{Rphi2omega}).

%%%%%%%%%%%%%%%%%%%%%%%%%%%%%%%%%%%%%%%%%%%%%%%%%%%%%%%%%%%%%%%%%%
\subsection{Matter and radiation domination\label{Rphi2:cosmology}}
%%%%%%%%%%%%%%%%%%%%%%%%%%%%%%%%%%%%%%%%%%%%%%%%%%%%%%%%%%%%%%%%%

The de Sitter solutions we found give us phenomenologically acceptable cosmic acceleration today.  Here we study the conditions under which these solutions can be extended to the early universe in the presence
of matter and radiation.  More concretely we seek a solution of the form
$\phi = \phi_0 + \phi_1$ where $\phi_0$ is a constant chosen
to give us phenomenologically acceptable cosmic acceleration today and $\phi_1$ is
a small time-dependent perturbation.  It is by no means obvious that this is the most interesting dynamical solution, but such
a quasi-static solution for the scalar field, if it exists, provides a particularly
simple way to satisfy all observational constraints.

Applying the FRW ansatz to the equations of motion of the $R\phi^2$ model we find
\begin{eqnarray}
\label{Reqt}
&& -R = {2 \over M_b^2 - \xi\phi^2}\Big[ {1\over 2} (-\rho + 3 p)^{\rm matter} 
+ {1\over 2} (h \dot\phi^2 - 4 (\rho_{\rm vac} + J\phi)) \\ \nonumber
&& \quad \quad \quad - 3(\xi \phi (\ddot \phi + 3 H\dot\phi) + \xi\dot\phi^2) \Big]
\end{eqnarray}
where we have taken the trace of the Einstein equations (\ref{F(R)Einstein}).   The scalar equation of motion is
\begin{eqnarray}
\label{heqt}
h (\ddot \phi + 3H \dot\phi) + {1\over 2} h' \dot\phi^2 = - J - \xi \phi R\,.
\end{eqnarray}
The static de Sitter solution obeys
\begin{equation}
\label{static}
J + \xi \phi_0 R_0 = 0 \,, \quad R_0 = 4(\rho_{\rm vac} + J\phi_0)/(M_b^2 - \xi\phi_0^2)
\end{equation}
where $R_0 = 12 H_0^2$ is the curvature today and where we defined
\begin{eqnarray}
\label{eff}
\rho_{\rm vac}^{\rm eff} \equiv \rho_{\rm vac} + J\phi 
\,, \quad M_{\rm pl}^2 \equiv M_b^2 - \xi\phi_0^2\,.
\end{eqnarray}
Here $M_b$ is the bare Planck mass (the quantity we denoted $\mp$ previously).

Going back in time, the source term for the $\ddot\phi$ equation is
$J + \xi \phi R \approx \xi \phi R$ because $R \propto a^{-3}$ during matter domination
(we will discuss radiation domination below). Assuming $\phi_1 \ll \phi_0$ we therefore
have
\begin{eqnarray}
\label{Hda}
{H \over a^2} {d\over da} \left[a^4 H {d\phi_1\over da}\right] \sim -3 {\xi \over h} \phi_0 H^2
\end{eqnarray}
where we have used $\ddot\phi_1 + 3 H \dot\phi_1 =
(H/a^2) (d/da)[a^4 H d\phi/da]$ and $R = 3 H^2$ during matter domination.
Integrating we have
\begin{eqnarray}
\label{Hdaint}
\phi_1 = -3 {\xi \over h} \phi_0 \int {da \over a^4 H} \left[\int a^2 H da \right] + A \int {da \over a^4 H} + B
\end{eqnarray}
where $A$ and $B$ are dictated by initial conditions.
The constant $B$ can be absorbed into $\phi_0$. The mode associated with $A$ decays with time,
and one can reasonably argue that for a wide range of initial conditions this mode will be
subdominant at all times relevant for observation.
There remains the driven mode, given by
\begin{eqnarray}
\phi_1 = - 2 {\xi \over h} \phi_0\ {\rm ln}\ a\,.
\end{eqnarray}
Clearly one can make $\phi_1 \ll \phi_0$ by choosing $\xi/h \ll 1$. The weak dependence
on $a$ means it is easy to satisfy $\phi_1 \ll \phi_0$ for all observationally relevant scale factors.
However to be self-consistent we also need to check that
the perturbation does not dominate over the matter energy density and pressure
in (\ref{Reqt}).
Examining the various contributions from $\phi_1$ and its derivatives, the dominant
contributions are of order $(\xi^2/h) H^2 \phi_0^2$ (such as from $\dot\phi_1^2$),
and the self-consistency requirement is that this is much smaller than
$\rho^{\rm matter} \sim H^2 M_{\rm pl}^2$. 
In other words, we need to make sure that
$(\xi/h) \xi \phi_0^2 \ll M_{\rm pl}^2$. Recall that the phenomenologically
interesting static de Sitter solution has $M_b^2 \gg M_{\rm pl}^2$, which basically
means $\xi \phi_0^2 \sim M_b^2$ (see (\ref{eff})). Therefore
a solution for the scalar field that does not significantly modify the expansion
rate at early times requires
\begin{eqnarray}
\label{xihcondition}
{\xi \over h} \ll {M_{\rm pl}^2 \over M_b^2} \sim {\rho_{\rm vac}^{\rm eff}
\over \rho_{\rm vac}}\,.
\end{eqnarray}
This requirement is rather severe, but it happens to be about the same as
the requirement for stability of the static solution.

What happens if we go back even further, to the era of radiation domination?
An interesting observation is that due to its traceless stress tensor radiation
makes no contribution to $R$.  However this doesn't mean $R$ vanishes, due to the stress tensor
for $\phi$, so in fact a quasi-static scalar field would mean $R \sim R_0 + \delta R$ during radiation domination.
Then the $\phi$ equation of motion (\ref{heqt}) tells us
\begin{eqnarray}
\nonumber
&& \ddot \phi_1 + 3 H \dot\phi_1 = - {\xi \over h} R_0 \phi_1 - {\xi \over h} \phi_0 \delta R \\ \nonumber
&& \quad \quad \quad \quad 
= - {\xi \over h} R_0 \phi_1 + 2 {\xi \over h} R_0 \phi_1 {\xi \phi_0^2 \over M_{\rm pl}^2}
- 6 {\xi \over h} {\xi \phi_0^2 \over M_{\rm pl}^2} (\ddot \phi_1 + 3 H \dot\phi_1) \\
\label{phi1dd}
&&  \quad \quad \quad \quad \approx 2 {\xi \over h} {M_b^2 \over M_{\rm pl}^2} R_0 \phi_1 - 6 {\xi \over h} {M_b^2 \over
M_{\rm pl}^2} (\ddot\phi_1 + 3 H \dot\phi_1)
\end{eqnarray}
where we've used the fact that $J + \xi \phi_0 R_0 = 0$.

We're allowing $\phi_1$ to evolve on cosmological time-scales, so
$H \dot\phi_1 \sim H^2 \phi_1$ which is much larger than $(\xi/h) (M_b^2/M_{\rm pl}^2) R_0 \phi_1$.\footnote{From (\ref{xihcondition}) we have
$(\xi/h) (M_b^2/M_{\rm pl}^2) R_0 \phi_1 \ll R_0 \phi_1 \sim H_0^2 \phi_1$.}  This means we can
drop the first term on the right hand side of (\ref{phi1dd}).  Also given (\ref{xihcondition}) we can
drop the last term on the
right hand side of (\ref{phi1dd}).  So we're left with a very simple equation
\begin{eqnarray}
\ddot \phi_1 + 3 H \dot\phi_1 \sim 0
\end{eqnarray}
with solution
\begin{eqnarray}
\phi_1 = A/a + B\,.
\end{eqnarray}
$A$ and $B$ are determined by initial conditions.
$B$ can be absorbed into $\phi_0$, and $A$ can always be chosen to
be small enough to satisfy any self-consistency requirements one needs
to impose.
(Think of it this way: suppose $A/a$ takes some natural value
in the very early universe, then for the $a$'s of interest
it is likely that $A/a$ is very small.)
The upshot is that a quasi-static approximation for the scalar field is valid
during radiation domination as well.

In summary, the presence of matter and radiation is consistent with a quasi-static scalar field as long as
(\ref{xihcondition}) is satisfied.  A similar analysis with similar conclusions can be made for the Gauss-Bonnet model.

%%%%%%%%%%%%%%%%%%%%%%%%%%%%%%%%%%%%%%%%%%%%%%%%%%%%%%%%%%%%%%%%%%
\subsection{Radiative stability\label{Rphi2:radiative}}
%%%%%%%%%%%%%%%%%%%%%%%%%%%%%%%%%%%%%%%%%%%%%%%%%%%%%%%%%%%%%%%%%

So far we have treated the $F(R)$ models as classical field theories.
But one of the central mysteries of the cosmological constant
is whether it can be protected from large radiative corrections.  In
this section we study quantum effects.  To be concrete we will focus
on the $R\phi^2$ model, although similar results should hold in general
$F(R)$ models.

First let's consider the model in Jordan frame.
\[
S = \int d^4x \sqrt{-g} \left({1 \over 2 \kappa^2} R - {1 \over 2} (\partial \phi)^2
- {1 \over 2} \xi R \phi^2 + J \phi - \rv \right)
\]
A key observation is that $\xi = J = 0$ is an enhanced symmetry point,
where the theory is invariant under shifts $\phi \rightarrow \phi +
{\rm const}$.  This means radiative corrections to quantities like
$\xi$, $J$ and the scalar mass $m$ (set to zero above) have to be
proportional to the symmetry-breaking parameters $\xi$ and $J$
themselves.  To summarize the results of our analysis: radiative
corrections to these symmetry-breaking parameters are under control if
$\xi$ is sufficiently small, roughly $\vert \xi \vert < 10^{-120}$.
However this argument does {\em not} address radiative corrections to
the vacuum energy $\rv$ itself.  Indeed we will find that, from the
point of view of quantum corrections to $\rv$, the $R \phi^2$ model is
just as unstable as a bare cosmological constant.

We will perform the rest of our analysis in Einstein frame.
\[
S = \int d^4x \sqrt{-\widetilde{g}} \left({1 \over 2 (\ke)^2} \widetilde{R} - {1 \over 2} \widetilde{h}(\phi)
\widetilde{g}^{\mu\nu} \partial_\mu \phi \partial_\nu \phi - \widetilde{V}(\phi)\right)
\]
We will concentrate on quantum corrections to the scalar potential
$\widetilde{V}$.  Before studying our model, consider a scalar field
theory in flat space with a potential
\[
V_{\rm flat} = \rho + J \phi + {1 \over 2} m^2 \phi^2 + {1 \over 3} g \phi^3 + {1 \over 4} \lambda \phi^4\,.
\]
In this flat-space theory a one-loop vacuum bubble should make a
correction
\[
\raisebox{-0.6cm}{\includegraphics[height=1.5cm]{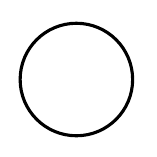}} \quad \Rightarrow \quad
\Delta \rho \sim \Lambda_{UV}^4
\]
where $\Lambda_{UV}$ is a UV cutoff scale.  A one-loop wart diagram
should make a correction to the scalar mass
\[
\raisebox{-0.6cm}{\includegraphics[height=1.5cm]{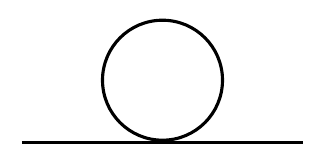}} \quad \Rightarrow \quad\Delta m^2 \sim \lambda \Lambda_{UV}^2\,.
\]
Finally a one-loop tadpole diagram should make a correction to the
source
\[
\raisebox{-0.6cm}{\includegraphics[height=1.5cm]{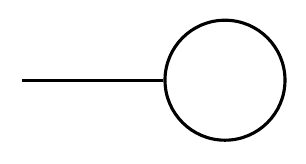}} \quad \Rightarrow \quad
\Delta J \sim g \Lambda_{UV}^2\,.
\]
This will generate a shift in the vev
\[
\Delta \langle \phi \rangle = - \Delta J / m^2
\]
which in turn produces a shift in the vacuum energy (for $J = 0$)
\[
\Delta \langle V \rangle = (\Delta J)^2 / 2 m^2\,.
\]

What do these results mean for the $R\phi^2$ model?  Setting $\phi =
\langle \phi \rangle + \delta \phi$, the effective potential of the
$R\phi^2$ model has an expansion in powers of $J \delta \phi /
\rve$.\footnote{Here $\rve$ is a fixed quantity, independent of
$\phi$.  It's useful to note that
\[
{J^2 \over (\rve)^2} = {4 \xi (\ke)^2 \rv \over \rve}\,.
\]}
\bea
\label{Vexpansion}
\widetilde{V}(\phi) & = & {\kappa^4 \over (\ke)^4} {\rv - J \phi \over (1 - \xi \kappa^2 \phi^2)^2} \\
\nonumber & = & \rve - {J^2 \over 4 \rve} (\delta \phi)^2 - {J^3 \over 4 (\rve)^2} (\delta\phi)^3
- {3 J^4 \over 16 (\rve)^3} (\delta \phi)^4 + {\cal O}(\delta \phi^5)
\eea
There are three classes of corrections we will study.

\goodbreak
\noindent
\underline{Corrections to $m_\phi^2$}

\noindent
In Einstein frame the scalar mass is\footnote{In Jordan frame the
scalar mass is much smaller, $m_{\rm Jordan}^2 \sim \xi H^2 \sim \xi
(\ke)^2 \rve \sim J^2 / \rv$.  This fits with our stability
analysis: given the Jordan-frame mass you might have thought $\xi =
{\cal O}(1)$ would be sufficient to guarantee stability on
cosmological timescales, but in Einstein frame you find that you
actually need $\xi \lesssim 10^{-120}$.}
\[
m_{\rm Einstein}^2 \sim J^2 / \rve\,.
\]
Given the quartic coupling in our model the radiative correction gives
\[
\Delta m_{\rm Einstein}^2 \sim J^4 \Lambda_{UV}^2 / (\rve)^3
\]
or
\[
{\Delta m_{\rm Einstein}^2 \over m_{\rm Einstein}^2} \sim {J^2 \Lambda_{UV}^2 \over (\rve)^2}
\sim {\xi \rv \over \rve} \cdot {\Lambda_{UV}^2 \over (\mpe)^2}\,.
\]
Radiative corrections to the scalar mass are stabilized for $\xi
\lesssim 10^{-120}$, assuming $\rv \sim (\mpe)^4$ and $\Lambda_{UV}
\sim \mpe$.

\goodbreak
\noindent
\underline{Linear source term}

\noindent
In Einstein frame we'll generate a linear source term for the scalar field
\[
\Delta J \sim J^3 \Lambda_{UV}^2 / (\rve)^2\,.
\]
This shifts the vev and in turn corrects the vacuum energy.
\[
\Delta \widetilde{V} \sim (\Delta J)^2 / m_{\rm Einstein}^2 \sim J^4 \Lambda_{UV}^4 / (\rve)^3
\]
The quantity we need to look at is
\[
{\Delta \widetilde{V} \over \rve} \sim {J^4 \Lambda_{UV}^4 \over (\rve)^4}
\sim \left({\xi \rv \over \rve}\right)^2 {\Lambda_{UV}^4 \over (\mpe)^4}\,.
\]
Again radiative corrections are stabilized for $\xi \lesssim
10^{-120}$, assuming $\rv \sim (\mpe)^4$ and $\Lambda_{UV} \sim \mpe$.

\goodbreak
\noindent
\underline{Corrections to $\rve$}

\noindent
Here the $R\phi^2$ model suffers from the usual instability of vacuum
energy with respect to radiative corrections: in Einstein frame,
no symmetry forbids graviton loops or $\phi$ loops from generating
\[
\Delta \rve \sim \Lambda_{UV}^4\,.
\]
This means the $R \phi^2$ model doesn't avoid the usual naturalness
problems associated with vacuum energy.  Although this is easiest to
see in Einstein frame, one can reach the same conclusion in Jordan
frame.  In Jordan frame graviton and $\phi$ loops, or loops of
standard model particle fields, should generate a shift $\Delta \rv
\sim \Lambda_{UV}^4$.  This change in $\rv$ can be compensated by
shifting $\langle \phi \rangle$ to preserve a small $\rve$, but the
necessary change in $\langle \phi \rangle$ completely destabilizes the
effective Planck mass.\footnote{The Gauss-Bonnet model gets around
this because a change in $\langle \phi \rangle$ does not affect the
Planck mass.}  To be more explicit about the difficulty, suppose we
have a correction in Jordan frame
\[
\Delta \rv \sim \Lambda_{UV}^4\,.
\]
You might think we've achieved something: with $\rv \sim
(\mpe)^4$ and $\Lambda_{UV} \sim \mpe$, don't we naturally have
$\Delta \rv / \rv = {\cal O}(1)$?  Unfortunately the relation
(\ref{Jtuning})
\[
{J^2 \mp^2 \over \xi \rv^2} = 1 + 10^{-120}
\]
requires $\Delta \rv / \rv \lesssim 10^{-120}$.  Equivalently, it
requires $\Delta \rv \lesssim \rve$.  So the $R \phi^2$ model is just
as sensitive to radiative corrections as a pure cosmological constant.

To summarize, with $\xi \lesssim 10^{-120}$ radiative corrections to
the non-constant parts of the scalar potential seem to be under
control.  It's curious that this value of $\xi$ was already required
for cosmological stability.  However the constant term in the
potential is not protected, and makes the $R\phi^2$ model just as
unstable with respect to quantum corrections as a bare cosmological
constant.  We expect this last conclusion to apply to any $F(R)$ model,
since any $F(R)$ model can be given a scalar-tensor description and
mapped to Einstein frame as in (\ref{stability1}).

%%%%%%%%%%%%%%%%%%%%%%%%%%%%%%%%%%%%%%%%%%%%%%%%%%%%%%%%%%%%%%%%%%%%%%%%%%%%%%%%%%%
\section{Further examples of $F(R)$\label{appendix:examples}}
%%%%%%%%%%%%%%%%%%%%%%%%%%%%%%%%%%%%%%%%%%%%%%%%%%%%%%%%%%%%%%%%%%%%%%%%%%%%%%%%%%%

In this appendix we present further examples of modified gravity
actions, built purely from the scalar curvature, which have
low-curvature de Sitter solutions.  As in section \ref{sect:F(R)} we
will write the action as
\[
S = \int d^4x \, \sqrt{-g} \, (\mpe)^4 F(y)
\]
where
\[
y = R/4(\mpe)^2\,.
\]
Recall that any function $F(y)$ satisfying (\ref{Fcond}), namely
\be
\label{Fcond2}
F(\epsilon) = \epsilon \qquad\quad
F'(\epsilon) = 2\,,
\ee
will have a low-curvature de Sitter solution with the right
effective Planck mass, where $\epsilon = 3 H_0^2 / (\mpe)^2 \sim 10^{-120}$.

Clearly an infinite number of functions
satisfy (\ref{Fcond2}): $F(y) = {\rm exp}\,(y^2 / \epsilon) - 1$,
$\,\,\sqrt{1 + 2y^2 / \epsilon} - 1$, $\,\,2 \epsilon \log(y/\epsilon)
+ \epsilon$, $\,\ldots$\ \ To be a bit more systematic about this we
make the ansatz
\be
F(y) = \sum_{n = -\infty}^{\infty} a_n y^n\,.
\ee
Then (\ref{Fcond2}) implies the conditions
\be
\sum_n a_n \epsilon^n = \epsilon, \qquad
\sum_n n a_n \epsilon^{n-1} = 2\,.
\ee
Here are a few examples.

{\bf Example 1.}  We can satisfy the necessary conditions with a single coupling
constant by taking $F(y) = y^2/\epsilon$.  That is, the gravity action
\be
S = \int d^4x \, \sqrt{-g} \, R^2 / 16 \epsilon
\ee
has a solution with both the correct Planck mass and the correct vacuum energy.  To
make the model compatible with solar tests we invert the Legendre transform and add
a large scalar kinetic term.
\be
S = \int d^4x \, \sqrt{-g} \, \left(\phi R - {1 \over 2} h (\partial \phi)^2 - 4 \epsilon \phi^2 \right)
\ee
One curious feature: neglecting the scalar kinetic term, this action is invariant under the scale transformation
\be
g_{\mu\nu} \rightarrow a g_{\mu\nu} \qquad \phi \rightarrow \phi / a\,.
\ee
This means the model has de Sitter solutions with arbitrary radius (and varying Planck mass).  It is an
extreme example of the fact that, although (\ref{Fcond2}) guarantees the existence of a solution with the
desired properties, it does not preclude other de Sitter solutions.

{\bf Example 2.}  With two coupling constants, the most obvious way to
satisfy (\ref{Fcond2}) is to take $F(y) = 2y - \epsilon$.  This just
corresponds to ordinary Einstein gravity with a small cosmological
constant.  Another option is to choose $F = -2 \epsilon^2 / y + 3
\epsilon$.  Going to the scalar-tensor description, and adding a large
scalar kinetic term, the latter example shows that an Einstein-Hilbert
term is not necessary to pass solar system tests.

\noindent
The examples we have considered so far do not interest us much,
because we want models with $a_0 \sim -1$ representing a Planck-scale
bare vacuum energy.  Our remaining examples have arbitrary $a_0$.

{\bf Example 3.} Choose
\be
a_0 = {\rm arbitrary} \quad , \quad
a_{1} = {3 \epsilon - a_0 \over 2\epsilon} \quad , \quad
a_{-1} = -\epsilon (a_0 + \epsilon) / 2
\ee
and the rest of the $a_n$'s vanish.  That is, the gravitational action is
\be
S = (\mpe)^4 \int d^4x \sqrt{-g} \left({a_1 R \over 4 (\mpe)^2} + {4 (\mpe)^2 a_{-1} \over R} + a_0\right)\,.
\ee
Inverting the Legendre transformation, and choosing the relation
between $f$ and $\phi$ to set $\phi = (\mpe)^3 / R$, this is
equivalent to the $R\phi^2$ model studied in section \ref{sect:Rphi2}.
The scalar-tensor action is
\bea
\nonumber
&& S = \int d^4 x \sqrt{-g} 
\left[\Big({a_1 (\mpe)^2 \over 4} - 4 a_{-1}\phi^2\Big) R  - {1\over 2} h 
(\partial\phi)^2 - V(\phi)\right] \\
&&V(\phi) = - 8 (\mpe)^3 a_{-1} \phi - (\mpe)^4 a_0
\eea
where we added a kinetic term for $\phi$.

{\bf Example 4.} Choose
\be
a_0 = {\rm arbitrary} \quad , \quad
a_{-1} = - 2 \epsilon (a_0 - 2 \epsilon) \quad , \quad
a_{-2} = \epsilon^2 (a_0 - 3 \epsilon)
\ee
and the rest of the $a_n$'s vanish.  That is, the gravitational action is
\be
S = (\mpe)^4 \int d^4x \sqrt{-g} \left({4 (\mpe)^2 a_{-1} \over R} + {16 (\mpe)^4 a_{-2} \over R^2} + a_0\right)\,.
\ee
This looks strange, but remember that for solar system tests one must
revert back to the scalar-tensor theory.  Inverting the Legendre
transform, and again taking $\phi = (\mpe)^3 / R$, we have the
equivalent scalar-tensor theory
\bea
\nonumber
&&S = \int d^4 x \sqrt{-g} 
\left[\Big(-4a_{-1}\phi^2 - {32a_{-2}\phi^3 \over \mpe}\Big)R
- {1\over 2} h (\partial\phi)^2 - V(\phi)\right] \\
&&V(\phi) = - 8 (\mpe)^3 a_{-1} \phi - 48 (\mpe)^2 a_{-2} \phi^2
- (\mpe)^4 a_0\,.
\eea

%%%%%%%%%%%%%%%%%%%%%%%%%%%%%%%%%%%%%%%%%%%%%%%%%%%%%%%%%%%%%%%%%%%%%%%%%%%%%%%%%%%
\section{More on Gauss-Bonnet\label{appendix:GB}}
%%%%%%%%%%%%%%%%%%%%%%%%%%%%%%%%%%%%%%%%%%%%%%%%%%%%%%%%%%%%%%%%%%%%%%%%%%%%%%%%%%%

In this appendix we study the classical stability of solutions to the
Gauss-Bonnet equations of motion, and show that for $\xi \lesssim 1$
the low-curvature de Sitter solution is metastable for more than a
Hubble time.

%%%%%%%%%%%%%%%%%%%%%%%%%%%%%%%%%%%%%%%%%%%%%%%%%%%%%%%%%%%%%%%%%%%%%%%%%%%%%%%%%%%
\subsection{Equations of motion\label{GB:eom}}
%%%%%%%%%%%%%%%%%%%%%%%%%%%%%%%%%%%%%%%%%%%%%%%%%%%%%%%%%%%%%%%%%%%%%%%%%%%%%%%%%%%

We consider a generalized Gauss-Bonnet action coupled to matter,
\be
\label{appendix:action}
S=\int d^4 x \sqrt{-g}\left[\frac{1}{2 \kappa^2}R-\frac{1}{2}h(\phi)\partial_\mu \phi
\partial^\mu \phi - V(\phi) - f(\phi) {\cal G}\right]+S_M
\ee
where the Euler density is
\be
{\cal G}=R^2-4 R^{\mu\nu}R_{\mu\nu}+R^{\mu\nu\kappa\lambda}R_{\mu\nu\kappa\lambda}\,.
\ee
Define the following tensor
\bea 
{\cal G}^{\mu\nu}&\equiv&{1\over \sqrt{-g}}{\delta\over\delta g_{\mu\nu}}\left[\sqrt{-g}f(\phi){\cal G}\right] \\
\nn &=& 2R\nabla^\mu \nabla^\nu f(\phi)+2 g^{\mu\nu}\left[2R_{\rho\sigma}\nabla^\rho \nabla^\sigma f(\phi)-R\Box f(\phi)\right] \\
\nn &&-8R^{\rho( \mu}\nabla^{\nu)} \nabla_{\rho} f(\phi)+4R^{\mu\nu}\Box f(\phi) -4R^{\mu\rho\nu\sigma}\nabla_\rho \nabla_\sigma f(\phi).
\eea
The equations of motion for the metric are then 
\be
\label{gequation}
R_{\mu\nu}-\frac{1}{2}Rg_{\mu\nu}=\kappa^2\Big[T^M_{\mu\nu}+T_{\mu\nu}^\phi-2{\cal G}_{\mu\nu}\Big]
\ee
where
\be
\label{Tphi}
T_{\mu\nu}^\phi=h(\phi)\left(\partial_\mu \phi \partial_\nu \phi-\frac{1}{2}g_{\mu\nu}(\partial \phi)^2 \right)-g_{\mu\nu}V(\phi)\,.
\ee
The equation of motion for the scalar field is
\be
\label{phiequation}
h(\phi)\Box \phi + {1\over 2} h'(\phi) \partial^\mu \phi \partial_\mu \phi -V'(\phi)-f'(\phi){\cal G}=0\,.
\ee
It's a remarkable property of Gauss-Bonnet gravity that one obtains
second-order equations of motion, even though one is starting from a
higher-derivative action which naively would be expected to generate
fourth-order equations \cite{Stelle:1977ry}.

In maximally symmetric spacetimes we have
\beas
&& R_{\mu\nu}-\frac{1}{2}R g_{\mu\nu}=-{1\over 4} R g_{\mu\nu}\\[2pt]
&& {\cal G}={R^2 / 6}\\
&&{\cal G}_{\mu\nu}=0\,.
\eeas
In the absence of matter, and assuming a maximally symmetric metric and
constant $\phi$, the equations of motion simplify to
\bea
\nn
&& R=4\kappa^2 V(\phi)\\
\label{vacuum2}
&& V'(\phi)=-\frac{1}{6}f'(\phi)R^2\,.
\eea
These are the Gauss-Bonnet analogs of (\ref{redundant3}),
(\ref{redundant4}) in $F(R)$ gravity.  We see that critical points
of the scalar potential are not what give us de Sitter solutions.
Rather we must look for points on the potential such that
\be
V'(\phi)=-\frac{8}{3}\kappa^4f'(\phi)V(\phi)^2\,.
\ee

%%%%%%%%%%%%%%%%%%%%%%%%%%%%%%%%%%%%%%%%%%%%%%%%%%%%%%%%%%%%%%%%%%%%%%%%%%%%%%%%%%%%%%%%
\subsection{Classical stability\label{GB:stability}}
%%%%%%%%%%%%%%%%%%%%%%%%%%%%%%%%%%%%%%%%%%%%%%%%%%%%%%%%%%%%%%%%%%%%%%%%%%%%%%%%%%%%%%%%

Here we show that for $\xi \lesssim 1$ the low-curvature de Sitter solution is
metastable on cosmological timescales.  Plugging in the FRW ansatz 
\be
\label{frw}
ds^2=-dt^2 + a^2(t) \vert d\vec{x} \vert^2\ ,\ \ \ \phi=\phi(t)
\ee
we obtain the Friedmann equation from the $\mu\nu=00$ component of
(\ref{gequation}),
\be
\label{friedmann1}
\frac{3}{\kappa^2}H^2 - \frac{h}{2}{\dot \phi}^2 - V(\phi) - 24 \dot 
\phi f'(\phi) H^3=\rho_M
\ee
where $\rho_M$ is the matter energy density.  Assuming constant $h$, the
scalar equation of motion (\ref{phiequation}) becomes
\be
\label{friedmann2}
h\left(\ddot\phi + 3H\dot\phi\right) + V'(\phi) + 24 f'(\phi)H^2
\left(\dot H  + H^2\right)=0\ .
\ee
In these equations
$H\equiv \dot a/a$ is the Hubble parameter and some useful identities are
\be
R= 6\dot H + 12 H^2,\ \ \  {\cal G}=24H^2(\dot H+H^2)\,.
\ee
The remaining components of (\ref{gequation}) do not yield independent
equations, but rather are consequences of (\ref{friedmann1}) and
(\ref{friedmann2}).

We are interested in fluctuations about a low curvature de Sitter
solution in which $H$ and $\phi$ are constant.  Setting
\be
H(t)=H_0+\delta H(t),\ \ \ \phi(t)=\phi_0+\delta\phi(t)
\ee
to first order in the fluctuations (\ref{friedmann1}) reads (in the absence of matter)
\be 
\label{flucmetric}
24 f' H_0^3\delta\dot\phi+V'\delta\phi-{6 H_0\over\kappa^2}\delta H=0
\ee
where $f'$ and $V'$ are evaluated on $\phi_0$.  Similarly (\ref{friedmann2}) reads
\be 
\label{flucscalar}
h\delta\ddot\phi+3hH_0\delta\dot\phi+\left(V''+24 f'' H_0^4\right)\delta\phi+24 f' H_0^2\delta\dot H
+96 f' H_0^3 \delta H=0\,.
\ee
Solving (\ref{flucmetric}) for $\delta H$ and plugging into (\ref{flucscalar}) gives
an equation for $\delta\phi$,
\be
\label{flucteqn}
\delta\ddot\phi+3H_0\delta\dot\phi
+\frac{V''-384 \kappa ^2f'^2 H_0^6+24 f'' H_0^4}{h+96 \kappa ^2 f'^2 H_0^4}\delta\phi=0
\ee
where we used (\ref{vacuum2}) to eliminate $V'$.
Specializing to our model we set
\be
V(\phi) = \rv - J\phi\,, \qquad f(\phi) = {1 \over 2} \xi \kappa^2 \phi^2\,.
\ee
Then (\ref{flucteqn}) reduces to (for $\rv \gg \rve$, and using (\ref{GBvev}) to eliminate the vev)
\be
\delta\ddot\phi+3H_0\delta\dot\phi - 16 {\xi \over h} H_0^2 \left({\kappa^4 \rv \over 1 + 4 (\xi/h)
\kappa^4 \rv}
\right) \delta\phi = 0\,.
\ee
This means the low-curvature de Sitter solution is unstable.  Assuming
$\rv \sim \mp^4$ and $\xi/h \lesssim 1$ the factor in parenthesis is
${\cal O}(1)$.  So the timescale for the instability is
\be
\tau \sim \left({h \over \xi}\right)^{1/2} H_0^{-1}\,.
\ee
Provided $\xi/h \lesssim 1$ the solution will be metastable for more
than a Hubble time, and Hubble friction will play an important role in
the evolution of $\phi$.

%\bibliographystyle{brownphys}
%\bibliography{lambda}
\providecommand{\href}[2]{#2}\begingroup\raggedright\endgroup

\end{document}